\documentclass{article}

\usepackage{arxiv}

\usepackage[utf8]{inputenc} 
\usepackage[T1]{fontenc}    
\usepackage{hyperref}       
\usepackage{url}            
\usepackage{booktabs}       
\usepackage{amsfonts}       
\usepackage{nicefrac}       
\usepackage{microtype}      
\usepackage{lipsum}
\usepackage{graphicx}
\usepackage{pgfplots}
\pgfplotsset{compat=1.8}
\usepackage{graphicx}
\usepackage{xcolor}
\usepackage{pdfpages}
\usepackage{listings}
\usepackage{enumitem}
\graphicspath{ {./images/} }

\title{Towards Explainable Social Agent Authoring tools: A case study on FAtiMA-Toolkit}

\author{Manuel Guimar\~{a}es \\
manuel.m.guimaraes@tecnico.ulisboa.pt \\
INESC-ID \\ Instituto Superior T\'{e}cnico, Universidade de Lisboa \\
0002-4538-1009
  \And 
Joana Campos \\
joanacfcampos@gmail.com \\
INESC-ID \\ Instituto Superior T\'{e}cnico, Universidade de Lisboa \\
  \And 
Pedro A. Santos \\
pedro.santos@tecnico.ulisboa.pt \\
INESC-ID \\ Instituto Superior T\'{e}cnico, Universidade de Lisboa \\
  \And 
Jo\~{a}o Dias \\
joao.dias@gaips.inesc-id.pt \\
Faculty of Science and Technology, University of Algarve, \\ CCMAR and INESC-ID \\
 \And
Rui Prada \\
rui.prada@tecnico.ulisboa.pt \\
INESC-ID \\ Instituto Superior T\'{e}cnico, Universidade de Lisboa \\
}

\begin{document}
\maketitle
\begin{abstract}
\textbf{Disclaimer: At the moment this article is currently awaiting submission review. Since this process usually takes a lot of time, feel free to use it this version as a reference until it is published. Thank you for reading, if you have any feedback please send us a message. Enjoy.}

The deployment of Socially Intelligent Agents (SIAs) in learning environments has proven to have several advantages in different areas of application. Social Agent Authoring Tools allow scenario designers to create tailored experiences with high control over SIAs behaviour, however, on the flip side, this comes at a cost as the complexity of the scenarios and its authoring can become overbearing.

In this paper we introduce the concept of \textit{Explainable Social Agent Authoring Tools} with the goal of analysing if authoring tools for social agents are understandable and interpretable. 

To this end we examine whether an authoring tool, FAtiMA-Toolkit, is understandable and its authoring steps interpretable, from the point-of-view of the author. We conducted two user studies to quantitatively assess the \textit{Interpretability}, \textit{Comprehensibility} and \textit{Transparency} of FAtiMA-toolkit from the perspective of a scenario designer. 
One of the key findings is the fact that FAtiMA-Toolkit's conceptual model is, in general, understandable, however the emotional-based concepts were not as easily understood and used by the authors. Although there are some positive aspects regarding the explainability of FAtiMA-Toolkit, there is still progress to be made to achieve a fully explainable social agent authoring tool. We provide a set of key concepts and possible solutions that can guide developers to build such tools. 
\end{abstract}


Sectionion{Introduction}
Socially Intelligent Agents (SIAs) have an ever increasing range of applications from conversational interfaces on websites to tutors or teammates in educational environments \cite{pecune2016evaluating, christoffersen2002make}, where they are equipped with tools to conduct human-like interactions. Amongst the most promising applications of SIAs are serious games and social skills training environments. In these virtual environments SIAs behaviours can range from reactive wandering in the background of a scenario to complex social interactions that provide social support or assist the player in some skill training \cite{pereira2014non}. These autonomous agents sense the environment and act intelligently and independently from the user, allowing them to train and adapt specific verbal and nonverbal behaviors in socially challenging situations \cite{bosman2018virtual}. 

Agent-based frameworks allow to simulate agent's cognitive and affective processes \cite{dias2005feeling, hartholt_all_2013, popescu2014gamygdala, marsella2009ema}, in such environments, and can produce intelligent and emotional behaviour, in an unbounded number of situations. Yet, it is up to the author of a scenario \--- typically instructors, therapists, or researchers \--- to manually describe how individual traits, goals, beliefs and actions interact, and guarantee character adaptability and consistency as events unfold. This includes defining a plot, writing rules of behaviour, creating dialogues, keeping track of the possible outcomes, among other things. While this can be manageable in narrow domains of application, for a serious game or social skills training content designer, using an intelligent agent framework can quickly become an overbearing task. As a response to those difficulties, data-driven approaches to automatic content and agent creation have become very attractive and are pursued widely in academic research and industry \cite{elnasr2020}. Yet, these approaches require large datasets tailored to the domain\footnote{As a result, they struggle in generating content for open-ended worlds} and offer little control to the scenario designer.

The main advantage of agent-based authoring tools is that it allows a scenario designer to have high control over content creation and target specific learning needs. Although previous research survey a set of challenges associated with the complexity and accessibility of the these tools \cite{westera2020artificial, spierling2009authoring}, we argue that tools scaffolded by understandable meta-models and metaphors will empower scenario designers and make them trust that they can easily create complex SIA interactions. Additionally, the way the author and the tool communicate should help the designer understand what comes next in the authoring workflow, and help them achieve their authoring goals. We refer to tools governed by these principles as \textit{Explainable Social Agent Authoring Tools}.

Explainability of social agent authoring tools not only is the extent to which a strong conceptual model can be easily understandable by authors, but also the extent to which a set of mechanisms can help the author, through interaction, explain the cause and effect of their actions. 
\cite{clancey2021methods} explored this view of \textit{explanation-as-interaction} making a parallel with Intelligent Tutoring System (ITS) design. Their view supports that XAI\footnote{Explainable Artificial Intelligence} methods should drop the assumption that providing an explanation consists is summarizing a complex process in a few lines of text or simply showing a graphic illustrating the reasoning process of an algorithm\footnote{These methods are usually referred as \textit{post-hoc explainability techniques} in the literature}. They advocate that the tools should promote understanding by interaction, by applying similar methods to those applied in ITS research: highlight important concepts, integration of fragmentary knowledge, reflect on previous experience, etc. 
We argue that understanding how an AI framework works requires encoding knowledge explicitly in a framework of knowledge (meta-model) and that will allow to create \textit{context-aware} authoring experiences that promote communication between the system and the user\footnote{in the form of authoring assistant, for instance.}. That implies design for explainability \cite{preece2018asking}.

The contributions of this paper are fourfold. First, we introduce the concept of \textit{Explainable Social Agent Authoring Tools} and we frame it within explainable AI (XAI) literature. Second, we examine whether an authoring tool, based on theoretical concepts is understandable and its authoring steps interpretable. Although there is an established assumption that theory-based architectures are understandable, we investigate if this is true from the point-of-the-view of the user, which has not been done before. We present data from two user studies, where we investigate whether FAtiMA-toolkit, which is grounded on Dennett's Intentional Stance \cite{dennett1987intentional}, is understandable. Third, data shows that while some artefacts are understandable other underlying concepts, namely emotions, that are not as easily understood. Furthermore, users choose between design long interactions or create scenario ramifications, due to the resulting complexity. Finally, we draw from the data and present a set of suggestions for leveraging machine learning approaches to drive explanation-as-interaction in agent-based authoring tools. It is our stance, that an \textit{interactive hybrid approach} to authoring is necessary to make it a smooth process that users recognize as trustworthy.

Sectionion{Approaches to Create Socially Intelligent Agents}
\label{sec:relatedwork}

The design rational behind Socially Intelligent Agents (SIAs) represent decades of work across different fields such as Social Sciences, Cognitive Science and Human Computer Interaction \cite{rampioni2021embodied}. Currently, there are two main approaches used to create SIAs: \textit{theory-driven} and \textit{data-driven}, both come with advantages and disadvantages. The former is a top-down approach that consists in developing computational models grounded on theoretical principles from the social sciences literature. The latter is  bottom-up  approach  where  the  behaviour  of the agents  is  generated/created from a large collection of examples of humans reacting and acting in different contexts. A less active line of research is the use of ontologies to support the decision-making of SIAs. Yet, they have gained a new momentum with newly data-driven models capable of making commonsense inferences about entities and events \cite{Hwang2021COMETATOMIC2O}. In this section, we describe characteristics of the tools under these two umbrella term and we provide a critical analysis these two approaches reflect \textit{transparency}, \textit{complexity} and \textit{interpretability} the core concepts of an understandable tool (refer to Section \ref{sec:eat} for more details). Furthermore, we also analyse the systems in terms of \textit{control} and \textit{auditability} as these concepts are central to social agents authoring tools.

\subsection{Theory Driven Agent Modelling Tools}

Theory-driven architectures are based on the premise that for creating realistic models for Intelligent Agents, we should look at how humans behave and try to better understand the reason behind our decisions. This line of thinking resulted in the rise of cognitive architectures that intend to capture, at the computational level, intelligent behavior using the underlying mechanisms of human cognition. EMA \cite{Gratch04,Marsella06,Gratch03} and FAtiMA (FearNot! Affective Mind Architecture) \cite{dias2005feeling} are examples of such architectures that have shown to be able to realize intelligent behaviours in different contexts.  EMA is an appraisal-based computational model that follows Smith and Lazarus's cognitive-motivational-emotive psychological theory \cite{lazarus1991emotion} , and extends SOAR \cite{Laird2012} by incorporating emotions. In EMA, the appraisal and coping mechanisms are built on causal representations developed for decision planning. Considering an example where a bird suddenly enters the room with an agent, EMA analyses the impact of this event in its current plans and goals. If there is a plausible hypothesis where the bird attacks and injures the agent it creates a threat to an existing goal (being healthy), which will be considered undesirable and will then generate an emotion. FAtiMA is an Agent-centered Architecture with planning capabilities designed to use emotions and personality to influence the agent’s behaviour. It was initially developed in 2005 for the purpose of driving the behavior of autonomous 3D characters in a serious game about bullying \cite{dias2005feeling}. At the time, its distinguishing feature was the OCC Model of Emotions \cite{ortony1990cognitive} that affected the character's planning and behavioral reactions \cite{dias2014fatima}.

The lessons learned in the design of the first emotional-social-agent architectures were passed on to recent generations of agent-modelling tools, namely, how the complexity and accessibility have restrained their widespread adoption \cite{westera2020artificial}. Ensemble \cite{treanor2016framework}, an iteration of the ``Comme il Faut'' architecture \cite{mccoy2010authoring} perfectly illustrates this.  In Ensemble, each agent has deep representation of the social state of the environment as well as the social models of other characters. One distinct feature of this framework is that, in order to make decisions, agents use a mechanism driven by both the structure of the social practice as well as a model of socio-cultural norms of the world. Moreover, the game-developer-oriented design led to the creation of not only its own authoring tool but it also a  representation system which provides both structure and a visual metaphor to the authoring process \cite{treanor2016framework}.

FAtiMA Toolkit\footnote{An updated version of FAtiMA (FearNot! Affective Mind Architecture)} is a collection of open-source tools designed to facilitate the creation and use of cognitive agents with social and emotional skills \cite{guimaraes2019accessible}. The toolkit facilitates the inclusion of a dynamic model of emotions deriving how the character looks and acts and how the player’s responses are evaluated. One distinct feature is that the framework it follows a character-centred approach rather than a plot-centred approach. The authoring is focused on defining general profiles (a set of rules) of how characters should respond emotionally in their games across different scenarios and contexts. The main advantage of this approach is that the characters’ behaviours are consistent across different contexts and no elaborate hard-coding is needed \cite{mascarenhas2021fatima}. 

\begin{figure}[ht]
\centering
\includegraphics[width=0.9\columnwidth]{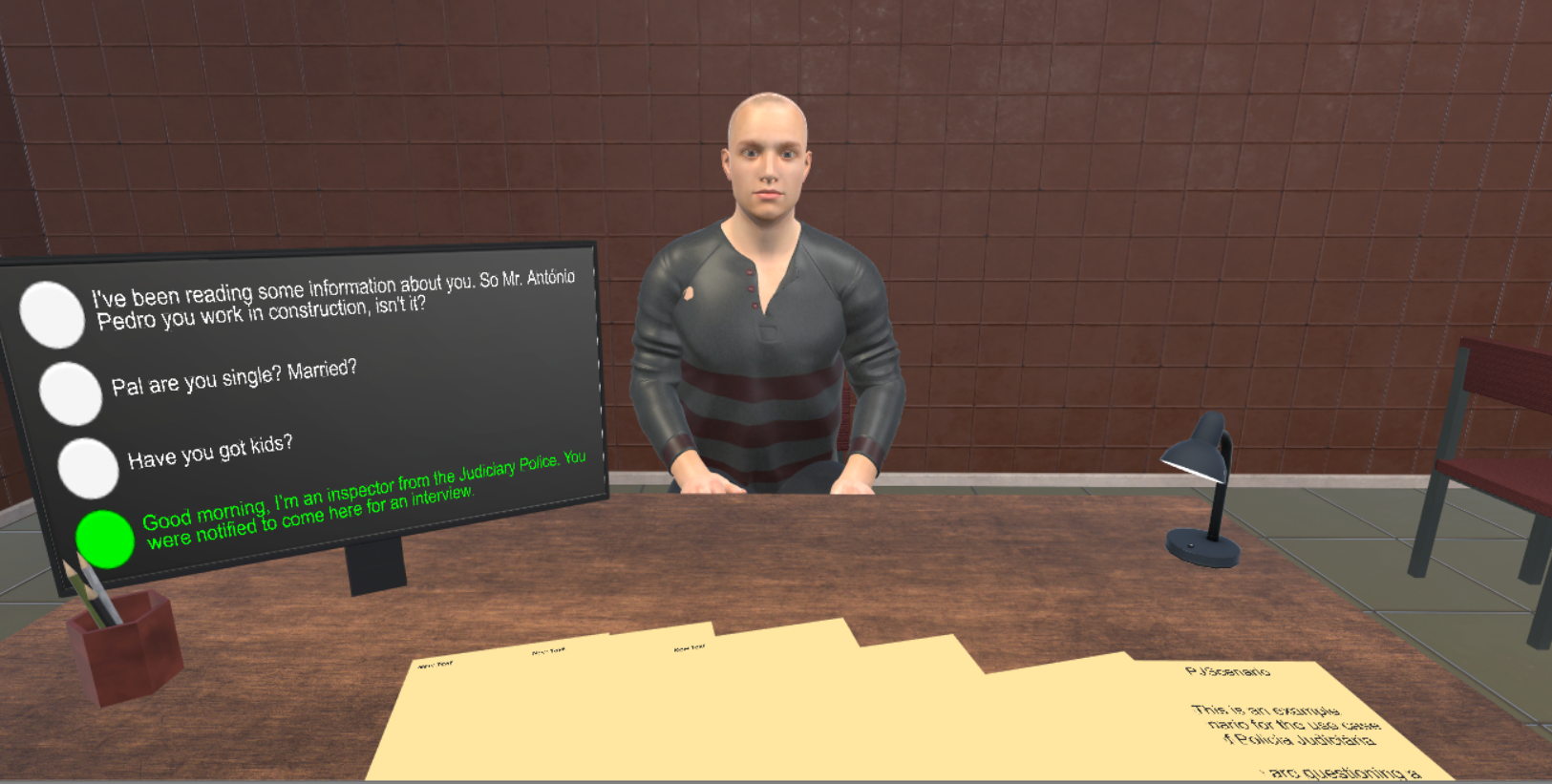} 
\caption{Example of a FAtiMA-Toolkit generated Virtual Agent \cite{guimaraes2020impact}}
\label{fig:vrdemo}
\end{figure}

At the other end of the spectrum, GAMYGDALA \cite{popescu2014gamygdala}, a computational model of emotions that is based on the OCC theory \cite{clore2013psychological} is presented as a , ``black-box AI independent, easily plugable emotion engine''. GAMYDGALA implements an appraisal mechanism that is independent from the how the characters reason. Essentially, developers need only to provide a list of goals for each agent and then specify which events will block or facilitate each goal. Based on that information, the engine will determine the changes on the character's emotional state.

Architectures developed under the ``theory-driven'' umbrella give \textbf{total control} to the user\footnote{Scenario author/designer} to create a myriad of \textbf{transparent} scenarios with infinite possibilities with a \textbf{high level of complexity}. Such complexity can be achieved by describing how the concepts (goals, intentions, actions, emotions, culture, etc.) that make the scaffolded of these computational artefacts interact in multi-agent scenarios overtime. Nevertheless, the ability of generating rich and plausible action sequences, \textbf{fully auditable}, comes at a cost. 

The theoretical basis required for authoring, makes the authoring process of agent modelling tools, particularly to users outside of the field, a strenuous task. The complexity of the authoring process of Ensemble led to DeKerlegand et al. \cite{dekerlegand2021pedagogical} work where a tutorial game to train users in the authoring of content was introduced.

Moreover, the task is heavily dependent on the designer’s ability to imagine a wide array of social situations and use their intuition to specify social behaviours and decision rules for the agent, which may be difficult to articulate \cite{liu2016data}. In addition to this, cognitive concepts such as goals and beliefs, which are quite familiar for AI researchers but not necessarily so for developers, which weights in on the decision of adopting such type of frameworks both in industry and academia \cite{mascarenhas2018virtual}, 
despite the empirical evidence of the benefits of using SIAs \cite{johnson2016face}.

\subsection{Data-Driven Approach for modelling Social Behaviour}
Apace with recent advances in deep learning, data-driven approaches have increasingly gained interest for motion generation (e. g. head motions, gestures, eye gaze, navigation within a crowd, etc.) for SIAs in a wide range of social scenarios. Their success in these lines of action have also led researchers to derive techniques that leverage examples of human-human interactions to acquire agents, both virtual and robotic, social behaviour. For instance, \cite{liu2016data} captured behavior elements, such as utterances, social situations, and transition rules from a large number of real human-human interactions. Using such data, not only did they show that were able to train a robot to answer questions, but also to proactively assist the human interlocutor in the task at hand. \cite{pecune2018field} deployed a SARA (Socially-Aware Virtual Assistant) in a conference site, showing that its data powered social reasoner was able to estimate the rapport level with the user and adapt the interaction accordingly.

Despite the success of the aforementioned approaches, generating plausible and varied social behaviour from existing data sources it is not as easy to experiment as in other domains, due to the lack of data to power AI-learning algorithms. To cope with the \textit{content generation challenge} in agent-based modeling of human social interaction \cite{feng2019exploring}, researchers explored different crowdsourcing techniques to gather social knowledge.
Using, mainly, Amazon Mechanical Turk as a recruitment platform, workers are tasked with creating social exemplar datasets such as alibi stories \cite{li2013story} or small narratives \cite{harrison2016learning}. \cite{feng2019exploring} tries to diversify the generated dataset by integrating an improvisational theatre training technique with its crowdsourcing task. The authors found that using two writers increased the diversity and the coherency of the stories. Although this line of research looks promising translating these elements to agent-based architectures, which offer control over the generated scenario, is not a straightforward task.

As highlighted above agent modelling architectures rely on a type of authoring that is oriented towards cognitive concepts such as goals, beliefs and emotions, among others \cite{mascarenhas2018virtual}. An option to translating natural language stories or descriptions into agent-readable concepts is to leverage recent developments in Natural Language Processing (NLP) and Machine Learning (ML) fields to a partial automatization of this task. This approach has been quite used in other fields, for example, in AI Planning. Much like in the Intelligent Virtual Agents field, there is the underlying assumption that users can formulate the problem using some formal language \cite{lindsay2017framer}. Here, knowledge acquisition tools have been used to extract the domain model through NLP. Framer uses Natural Language descriptions, written by users, as input and is able to learn planning domain models \cite{feng2018extracting}. In a similar vein, \cite{janghorbani2019domain} introduced an authoring assistant tool to automate the process of domain generation from natural language description of virtual characters with the objective of bridging the authoring and the planning tasks.

The most sought output using data has been to generate narrative plots, using story-based datasets \cite{tambwekar2018controllable}. For instance \cite{ammanabrolu2020automated} developed an automated storytelling tool to order the plot of stories using causal and commonsense knowledge \cite{ammanabrolu20world}, using COMET \cite{bosselut2019comet}, a transformer-based language model designed for commonsense inference.



While data-driven techniques might hold the key to easing the authoring burden \cite{elnasr2020} they are not without its issues. 
 Crowdsourcing, despite its usefulness as a means of creating a focused and task-oriented dataset, produces a lot of noisy and low-quality data \cite{becker2019crowd}. Although researchers have tried to minimize these effects, finding the most efficient way to collect training data via crowdsourcing still remains an open question, \cite{liu2019optimizing}. 
 
 In addition to this, data-driven approaches all face the same potential pitfall as other data-heavy systems   \cite{ammanabrolu2020automated}. Namely, they are prone to echoing the biases present in  \cite{sheng2019woman}, limiting their ability to produce generic and independent results. Moreover, deep learning algorithms, quite used in data-centric approaches, are a class of Machine Learning algorithms which tend to \textbf{sacrifice transparency} and \textbf{interpretability} for prediction accuracy. The intricacy of these methods, makes it so that the outputted model is inherently uninterpretable to human users \cite{rai2020explainable}. For that reason models of behaviour purely data-driven \textbf{deny control} to the user and sequences of behaviours are \textbf{not auditable}. In learning scenarios, not only is important to understand the learned model but also to have control over it and, if needed change and tune the agent's learning process \cite{eberhart2014neural}.

\subsection{Ontology-Based Architectures}
\label{sec:ontologies}


Ontologies provide the terms to describe a domain and the semantics associated with them. In addition, ontologies are often accompanied by logical rules that constrain the meaning assigned to those terms. These constraints are represented by inference rules that can be used by agents to perform reasoning. For SIAs the ability to reason about their surrounding environment is key.




In order to help agents understanding the world they are in, \cite{balint2013macgyver} used hierarchies of ontologies for both actions and objects. The Decision Making process combined with affordance theory \cite{gibson2014ecological} is then able to find viable objects for an action. Regarding interaction, KRISTINA \cite{wanner2017kristina} uses an advanced ontology-based knowledge model that is capable of capturing the content of the multimodal interactions during a conversation. Flexible dialogue strategies are driven by a knowledge representation including ontologies with relevant background information collected from the web. This system highlights that ontologies are typically developed with a pragmatic focus, having in mind a context and an intended use for a particular domain, generally being developed following a design method or methodology \cite{bermejo2011engineering}. Finally, the agents that are quite capable of making inferences regarding their environment and even solve social tasks it seems that during the development of these technologies the authoring experience was left as an afterthought.



\subsection{General Comments}
Regardless of the approach, the adoption of agent-based frameworks both in industry and academia is highly connected to the authoring experience of agent-modelling tools. Most of the work in the field of HCI, some of which described in this section, is focused on the impact and quality of the agents themselves. \textit{Are the agents believable?} \textit{Do they generate engagement?} \textit{Is there a positive or negative impact in human-agent interaction?} However, not much has been said in terms of the effort necessary to create rich environments and the link to their (or lack of) adoption.

Approaches to social agent modelling rooted on cognitive concepts (e.g EMA, PsychSIM and FAtiMA) or social physics (e.g. Emsemble) are by definition quite strong in terms of understandability and interpretability. Rule-based systems traditionally used are inherently transparent by themselves and also with a natural and seamless relation to human behaviour. Terms such as beliefs, goals and intentions are the core of agent-based tools, by improving their \textbf{interpretability} can contribute to lowering the entry barrier for inexperienced authors. On the other hand, one major issue with theory driven approaches is their general lack \textbf{scalability}. Complex interactive scenarios with multiple characters and moving pieces can lead to designers quickly finding themselves unable to handle the large authorial effort. As we have also mentioned in Section \ref{sec:relatedwork} developments in Natural Language Processing (NLP) and Machine Learning fields have led to the automatization, or at least partial automatization, of this task.  Authors are asked to provide textual input such as stories or scrips and the system is able to compute and transform it into intelligent character's scenarios \cite{harrison2016towards, li2014data}. In order to create trustworthy and more believable intelligent agents we need better tools to assist their design. We refer to these tools as \textit{Explainable Social Agent Authoring Tools}.

Sectionion{Explainable Social Agent Authoring Tools: Why and How}
\label{sec:eat}
Generation of explanations is a key issue in AI since the development of expert systems. In the earliest AI systems there was a need to make them more \textit{transparent} by providing traces of the reasoning performed by a complex set of rules \cite{preece2018asking}. Deep machine learning models and their opacity renewed interest in Explainable Artificial Intelligence (XAI), which attempts to expose in an understandable manner algorithms' mechanisms and predictions \cite{doran2017does}. Although theory-driven authoring tools stand comparison with expert systems, the new research in XAI has opened the discussion about explainability and is making an effort to establish a common framework for development and analysis. For those reasons we  use the terminology of this broader view of XAI.

\cite{arrieta2020explainable} surveyed that the motivations for XAI are different across audiences, but can be clustered around the following goals: trustworthiness, causality, transferability, informativeness, confidence, fairness, accessibility, interactivity and privacy awareness. Making a parallel with previous research, the motivations behind an \textit{Explainable Social Agent Authoring Tool} are \textit{accessibility} and \textit{interactivity}, which describe what such tool should compel. While the \textbf{accessibility} of authoring tools can be improved by creating better user interfaces, the issues that prevent the wide adoption of these tools go beyond interface usability. Previous research suggests that understandable conceptual models should guide the procedural principles of authoring tools \cite{spierling2009authoring, mascarenhas2018virtual}. An explainable tool, scaffolded by theoretical concepts that the author can easily identify, will ease the authoring burden facilitating the adoption by technical and non-technical users. Such conceptual-model lays out the building blocks needed to explore \textit{explanation-as-interaction} proved successful in ITS research for understanding and learning. Hence the other main goal that support the need of an \textit{Explainable Social Agent Authoring Tool} is \textbf{interactivity}, because it is the way that the user and the system are able to communicate that make the success of the tool \cite{clancey2021methods, langley2017explainable}.

Asking for an explanation is to ask a \textit{why} question, expecting a \textit{because} answer. Hence, explanation is about understanding \cite{van2014explaining}. Understanding is the ability to make the connections between several pieces of knowledge and identify causal and relational patterns \cite{van2014explaining, keil2006explanation}. This definition is central to the thesis of this paper that states that understanding how an AI framework works requires encoding knowledge explicitly in a framework of knowledge (meta-model). That is, instead of devising methods to generate explanations design for explainability \cite{preece2018asking}. This design consideration has broader implications on understanding of an authoring tool (and its capabilities), as it facilitates the conception of different layers of knowledge such as identifying rules that only work under specific conditions, causal chains of actions or make connections between artefacts grounded on the building bones of the system. 



\begin{figure}[ht]
\centering
\includegraphics[width=0.9\columnwidth]{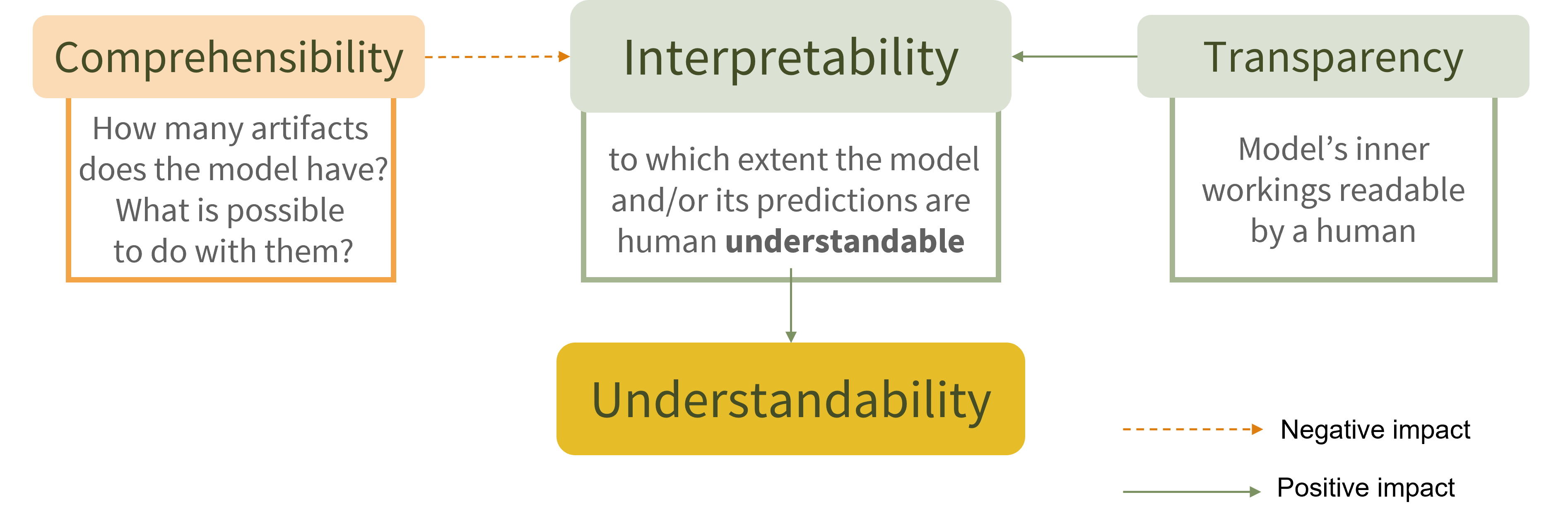} 
\caption{Understandability is the core concept in XAI. Understandability of a model is negatively influenced by \textit{Comprehensibility} (or complexity  of the model) and Positivey influenced by a model \textit{Transparency}.}
\label{fig:understandability-schema}
\end{figure}

From the analysis of the literature there are 3 main concepts that influence the explainability of a model, all grouped under an umbrella term \--- \textit{understandability}. \cite{arrieta2020explainable}'s analysis identified that understandability is the core concept of XAI and heavily influenced (positively and negatively) by a set of nested forces: Interpretability, Comprehensibility and Transparency (Figure \ref{fig:understandability-schema}). 
\textit{Interpretability} refers to the ability of the system being able to demonstrate that a given input \textit{i} will produce an output \textit{o} in a way that a human can understand.
\textit{Comprehensibility} is usually tied to the model complexity and one's ability to express information in natural language. In opposition to the other concepts, high complexity usually results in lower interpretability of a model. \textit{Transparency} can be promoted by showing how things work and can be divided in three different levels: simulatability, decomposability and algorithimic transparency. To better illustrate what that means consider a brief comparison between decision trees, rules and linear models, which are by design transparent models \cite{freitas2014comprehensible, huysmans2011empirical}. Decision trees can be viewed as transparent models due to their structure. It helps users to focus their analysis on the most relevant attributes and its hierarchical aspect provides information about the relative importance of different attributes \cite{freitas2014comprehensible}. While decision trees are widely adopted for their graphical representation, decision rules, on the other hand have a textual representation \cite{guidotti2018survey}. Textual representation does not instantaneously describe the most relevant components of rule. However, its interpretability can be increased through positional information. For instance, the conditions of a rule can be ordered by their importance. 


Rule-based approaches are the central method to represent knowledge in expert systems and it is the main mechanism for authoring in FAtiMA-toolkit. In theory it helps analysing and understanding the model and ultimately maximize control over the output. The set of characteristics make the tool by definition understandable. It is important to underline, however, that the user's cognitive ability and experience must be taken into account along with the interpretability and comprehensibility of the model in use. Certainly, different audiences will have different interpretations of the same model therefore understandability can be seen as a two-sided discussion: model understandability and human understandability \cite{arrieta2020explainable}. For that reason, in our work we explore to what extent a theory-based authoring tool is \textit{understandable} (see Section \ref{sec:evaluation}).

\begin{figure}[ht]
\centering
\includegraphics[width=0.8\columnwidth]{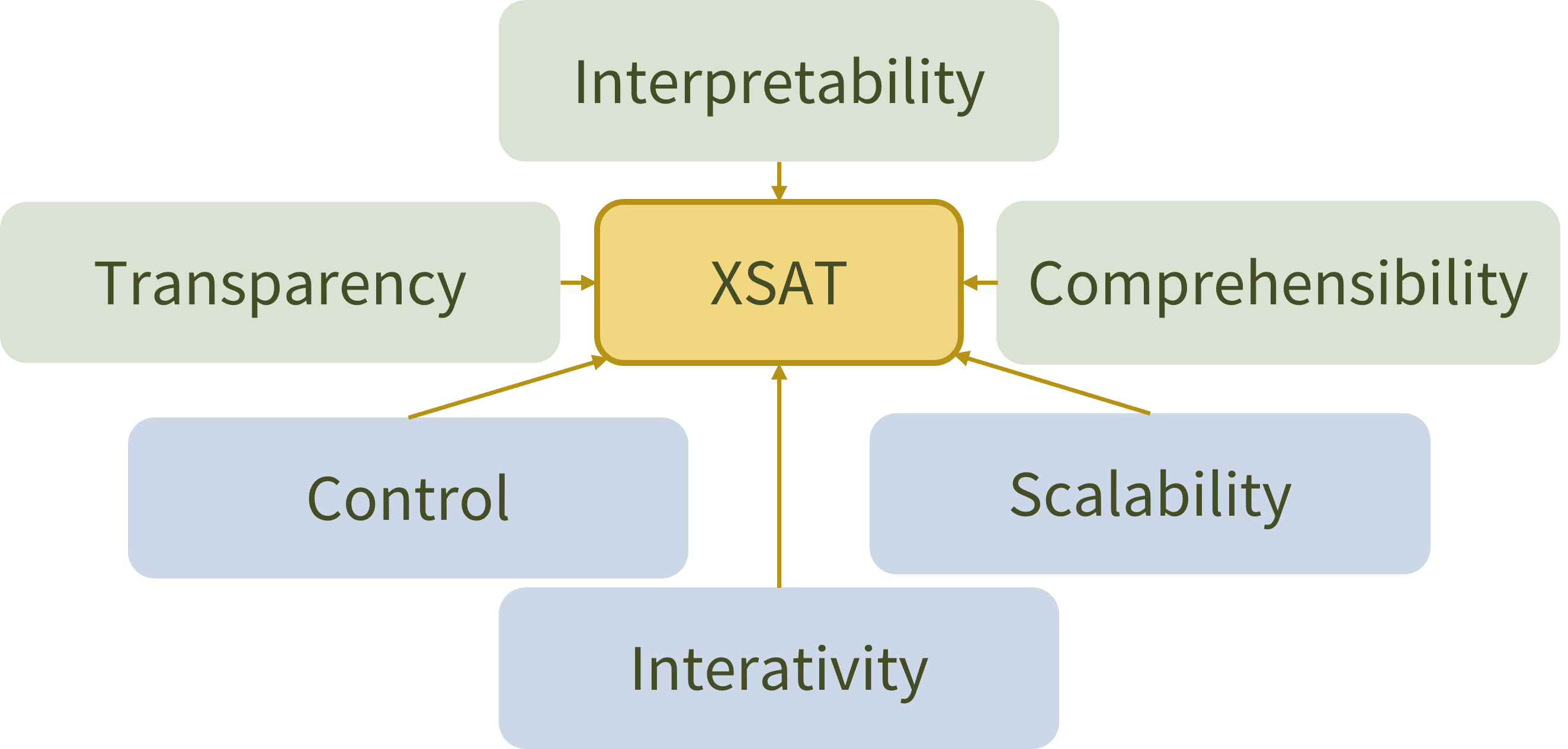}
\caption{Comprehensibility, Interpretability, Transparency, Control, Interactivity and Scalability are the forces that influence \textit{Explainable Social Agent Authoring Tools} (XSAT)}
\label{fig:xsat-forces}
\end{figure}

Given these points, three defining concepts of \textit{Explainable Social Agents Authoring Tools} (XSAT) are \textbf{comprehensibility}, \textbf{interpretability} and \textbf{transparency}, which are the key concepts when analysing AI algorithms through the lens of XAI. In addition, our analysis of the current state-of-the-art in tools for creating SIAs, in Section \ref{sec:relatedwork}, identified three other important forces that should guide the design of \textit{XSATs} and influence its degree of explainability (see Figure \ref{fig:xsat-forces}). Those are \textbf{control}, \textbf{scalability} and \textbf{interactivity}. \textit{Control} refers to the degree of possible customization in scenario definition that should go beyond basic properties definition. Systems like FAtiMA-toolkit allow authors to fully define a character (including believes, intentions and actions not restricted by pre-defined categories) and how characters interact (a walk-through on what consists a authoring task is presented in the next section). \textit{Scalability} concerns the ability of a system to handle an increasing number of characters in a scenario and their resultant interactions. This aspects tends to be negatively influenced by the amount of control present in an authoring tool or in the case where automatization is present, it will influence negatively \textit{Control} (in section Section \ref{sec:suggestions} we discuss how to balance these two forces). Finally, \textit{Interactivity} specifies the type (and amount) of communication possible between the user and the system. We envision that XSAT high in interactivity are able to help the user in (and not limited to) identify important concepts overlooked by the author, integrate fragmentary knowledge based on causal chains of actions, and reflect on the created artefacts and what it is possible to achieve with them.


Sectionion{Research Questions}
Following the previous sections we are introducing a new concept \--- \textit{Explainable Social Agent Authoring Tool}\--- that allows us to clearly define how explainable a social agents' authoring tool is or can be. This is of extreme relevance for the field as it directly linked to system adoption.

There is an established assumption that explainability of a system is only worth to study if those follow data-driven approaches. Theory-driven architectures are assumed to be understandable (  transparent, interpretable and comprehensible) based on the fact that they are built upon well established psychological theories. We consider, however, that it is desirable to understand whether current state-of-the-art tools are in fact understandable from the point-of-view of the user. A recent work by \cite{dekerlegand2021pedagogical} highlights the difficulties that authors (from different backgrounds) experience when confronted with supposedly easily identifiable concepts. The response was to create a tutorial to circumvent the issues that aroused during a highly complex authoring task. This underlines the importance of studying these aspects before moving to create tools that assist the user during authoring. It is important to understand how to design for explainability.

These considerations drive our research questions for the study, focused on understandability, presented later in the paper.

\begin{description}
\item[RQ1] Are the formalisms and resulting artefacts of Theory-Driven Social Agent Frameworks positively influencing interpretability and transparency of the tool?
\end{description}

This question intends to address whether the underlying concepts of a theory-driven architecture are understandable, by a user. 
As we shall see, most are  rooted on \cite{dennett1987intentional}'s Intentional Stance, who postulated that a rational agent with beliefs, desires and other mental states exhibit intentionallity and thus the agent's past and future behaviour is reliably predictable. He asserts that the use of the intentional strategy in our daily lives is so ``effortless that the role it plays on shaping our expectations can be easily overlooked'', but it is an ``extraordinarily powerful tool in prediction''. Emotions guide our daily decisions and help us to establish goals. 
Thus, the understandability of the computational models of emotion usually associated and used by agent-modelling frameworks should also be studied.

\begin{description}
\item[RQ2] How complex are the scenarios that a author can create using their knowledge about the underlying concepts that make up FAtiMA-toolkit?
\end{description}

Comprehensibility is normally tied to the evaluation of the model complexity. We wish to explore how complex are the scenarios an author can create after gaining some experience with the tool. As the users create particular instantiations of relations, actions, rules and
constraints for a particular domain the complexity increases and dealing with the ramifications can be more difficult. As such, this research question can be broken down
into more specific sub-questions:

\begin{enumerate}[label=\alph*)]
    \item Do the authors feel some artefacts are more familiar than others?
    \item What artefacts do the authors use to create rich socially interactive scenarios? 
\end{enumerate}

Before trying to answer these questions it is necessary to understand the artefacts behind traditional Theory-Driven approaches and the degree of complexity and work necessary to create affective human-agent interactions.

Sectionion{Authoring an Intelligent Agent Scenario}
\label{sec:authoring}

In this section we describe the steps involved in authoring SIAs with examples of how that is achieved in different theory-driven social agent architectures. The intent of this exercise is to provide a clearer picture of what the ``authoring process'' is and illustrate the amount of effort needed to create artefacts from different componenents. 

An intelligent agents follows a perception-action loop where agents perceive its environment through sensors and act upon it through actuators \cite{russell2002artificial}. Socially intelligent agents (SIAs) are agents that are able to connect and interact with humans that show aspects of human like social intelligence \cite{dautenShahn1998art}. The task of endowing agents with social behavior has many alternative paths, however, in general, all revolve around the same social science's concepts such as beliefs, goals, actions and emotions.

\subsection{Scenario Planning}

Regardless of the technology, platform or tool being used it is considered good practice to write a short description or story of what the desired interaction experience users are expected to have. This planning stage is characterized by the creation of some sort of outline of the experience and its interaction goals. In the case of Social Teaching Environments, the interaction goals are typically making sure participant are taught a particular topic, for instance, how to follow the standard police interview protocol \cite{guimaraes2020impact}.

Once the general outline of the interaction has been drawn the following step is to start its implementation.

\subsection{Beliefs, Desires and Goals}
Often, the first step when implementing an agent-based experience is the development of the agent's themselves including their beliefs, desires and goals. In general these are tied to the scenario's context, for instance, if the interaction is in a restaurant, it is logical for the agents to have the notions of ``food'' and ``hunger''. In EMA \cite{Gratch04,Marsella06,Gratch03}, like most theory-driven frameworks,
the state of the world is represented as a conjunction of propositions. Each belief corresponds to ``commitments to the truth'' and their value is binary (although there is support for the use of probabilities): ``Apple-Have|P=1.0'',  ``U-raised|P=1.0'', "Injured|P=0.0" represent beliefs where the actor has an apple, its umbrella is raised and it is uninjured.  Thus the representation of this state would correspond to the following conjunction:


\begin{figure}[ht]
\centering
\includegraphics[width=0.4\columnwidth]{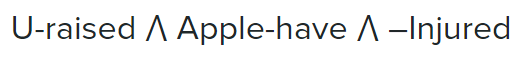} 
\label{fig:radar-chart}
\end{figure}

In general, there is no rigid convention for how beliefs and goals are declared as long as they are consistent across the scenario and its framework. Another example, contrasting to EMA, FAtiMA-Toolkit's beliefs can have any value, not just true or false. 

\begin{lstlisting}
Has(Apple)=5
Edible(Apple)=True
is(Hungry)=True
\end{lstlisting}

\subsection{Actions}

Actions typically have both (pre)conditions and effects. Consequently they tend to be represented as rules. If an action's conditions are verified then the agent might decide to perform that action. In addition to this, each action usually has an utility value associated, that represents its importance or intention. The result of the execution of an action is described by its effects. Some frameworks model the possibility of failure or it never being completely executed. For instance, the Social Practices of Comme il Faut, now known as Ensemble \cite{mccoy2014social}, allow for the target of the action to, for instance, reject an insult or accept a flirtatious invitation. 


The action space of each agent must be consistent with the context of the scenario. Once again, if the interaction is set in a restaurant while an agent might have actions related to being in a beach (such as playing with a ball) it wouldn't make sense to use them, nor not having any actions restaurant-related. Hence, the agent's action should use the beliefs and goals set in the previous stage of development.

An action rule defined in FAtiMA-Toolkit \cite{guimaraes2019accessible} uses as conditions the value of beliefs present in the agent's knowledge base. The subsequent action rule captures the following phenomenon: ``An Agent performs the Eat action towards [item] if the [item] is edible, if the agent is hungry and if it has at least 1 [item] in their knowledge base'':

\begin{lstlisting}
Action: Eat
Target: [item]
Priority: 1
Conditions:
  Edible([item]) = True
  Is(Hungry) = True
  Has([item]) > 0
\end{lstlisting}

Ensemble uses Social Practices that define a pattern of social interaction such as a ``Compliment'' or an ``Insult''. Each Social Practice is composed by stages with a selection of actions \cite{treanor2016framework}. As an illustration, an agent, ``Tom'' wants to perform a ``Sincere Compliment'' to another agent: ``Sarah''. ``Tom'' starts by, within the first stage of the ``Sincere Compliment'' Social Practice, to rank each action from the available action's list: ``friendly compliment'', ``romantic compliment'' etc...and then choosing the highest ranking one. The ranking is influenced by the agent's beliefs, goals and the world around it. 

Once ``Tom'' to has performed its action ``Sarah'' can respond and the Social Practice moves onto its second stage. In this example, ``Sarah'', according to its internal state, opts chooses the stage: ``Joking Response'' which has the following possible actions: ``joking about romantic'', ``joking about friendly'', ``joking about backhanded'', ``joking about flatter''. Once the choice is made the agent performs the action. For example if the ``joking romantic'' was chosen the response could be something as: ``Thanks, put it in writing, and I might consider it''. 

It is important to note that both of these frameworks, FAtiMA-Toolkit and Ensemble \cite{guimaraes2019accessible, treanor2016framework} have an Authoring Tool whose objective is to facilitate the author's work. It provides a user interface where authors can easily create, add, edit and remove different components within the scenario. Using these tools users avoid having to manually code everything, unlike their predecessors, \cite{dias2005feeling,mccoy2010authoring}). However, similar to the examples framed here, all of these different parameters; actions, conditions and dialogues need to be manually authored. 

Depending on the architecture, the result of each action may need to be explicitly authored. In FAtiMA-Toolkit, for instance, the World Model is responsible for generating different effects for different actions using, like most of its componenents, a rule-based approach \cite{mascarenhas2021fatima}. In order to maintain the immersiveness, the suspension of disbelief, during the interaction, authors tend to design action effects that are coherent with what happens in real-life but also serve the set upon interaction goals. Often this is achieved by prompting emotional responses by the agents. 

\subsection{Emotions}

There is a wide variety of theories of emotions which, in turn, generated a wider range of diverse computational models \cite{ortony1990cognitive, mehrabian1996pleasure, scherer1999appraisal}. agents that are able to express and incorporate affects into rational processes and react to event in the world \cite{pudane2017human}. 

GAMYGDALA \cite{popescu2014gamygdala} follows the OCC Model of Emotions \cite{ortony1990cognitive} and automatically generates emotions based on the agent's goals. Designer's only need to define which type of events are relevant to each goal. For example, for the popular videogame Pac-Man, the goals of the ghosts can be defined as ``get eaten'' with a negative utility, degree of importance, -1, and ``catch pacman'' with a positive utility of 0.7. If a ghost does get close to Pac-Man it will increase the likelihood that its goal will be satisfied causing the emotion of ``hope'' to be triggered. Alternatively if pac-man was in the "eating ghosts state" the belief of pac-man getting close would influence the ``get eaten'' goal therfore generating feelings of ``fear''.  
In order to use GAMYGDALA developers must define the goals and beliefs using the engine as follows:

\begin{lstlisting}
EmoBrain brain = new EmoBrain();
brain.Goal("get eaten", -1.0f);
brain.Goal("catch pacman", 0.7f);
brain.Belief("pacman is close", 1f, "ghost")
brain.AffectsGoal("pacman is close", "catch pacman", 1f);
brain.Update();
\end{lstlisting}

On the other hand, and, as we mentioned before, authoring tools release the burden of writing code from the interaction designers. However they still heavily rely on the author to create emotional rules and understand the emotional computational model behind the framework. Similar to GAMYGDALA, FAtiMA-Toolkit \cite{mascarenhas2021fatima} uses an appraisal system based on the OCC Model of emotions \cite{ortony1990cognitive}. Authors have the ability to write rules that capture generic or specific actions but must also decide which appraisal variables are triggered. For example, in the following case, the act of receiving food from the agent ``Waiter'' triggers the ``Desirability'' appraisal variable which according to the OCC model will generate feelings of Joy: 


\begin{lstlisting}
Subject: Waiter
Action: Give([food])
Target: John
Appraisal Variables:
  Desirability = 5
\end{lstlisting}

\subsection{Dialogue}

At its core, social interaction is essentially communicating with someone or with something. The most effective way of communicating with a SIA is through dialogue. In fact, while in the past most of the dialogue was manually authored, and part of it still is, its importance to the human computer interaction field has led to a wide range of different conversational systems, scripting tools and automatically generated content \cite{ma2020survey}. 


In order to tackle the challenge of creating and managing dialogue within a social agent simulation, Aljammaz et al. \cite{aljammaz2020scheherazade} integrated chatbots and knowledge model representation on top of the Ensemble engine \cite{treanor2016framework}. The chatbots are used as a layer of abstraction between the human interactor and the Emsemble Characters allowing open-text input to be taken into account by the scenario. For example, the player could interact with the agent by typing ``Hello John''. This would be recognized by the bots as the pattern it as the pattern “hello[*]” which maps to the Ensemble action ``Greet''. The engine would then calculate if the action succeeded  and, if so, perform the ``GreetSuccess'' action passing that information on to the rest of the system, namely the chatbots, generating the dialogue associated with that social exchange.

The solution adopted in FAtiMA Toolkit is to have an explicit dialogue tree structure without any logic. Instead, it merely informs agents of how many options they can choose from in a given dialogue state.
The logic of the dialogue is handled by a specific action called: ``Speak''. Through it the author is able to control the current and the next state of the conversation along with additional optional parameters. For example, ``An agent that chooses to speak with ``John'' using a `Rude' style when its mood is negative:

\begin{lstlisting}
Action: Speak([currentState],[nextState], - ,Rude)
Target: John
Priority: 2
Conditions:
  DialogueState(John) = [currentState]
  NextState(DialogueState(John) = [nextState]
  Mood(SELF) < 0
\end{lstlisting}

The conditions of each action are evaluated against the beliefs of each agent. Thus, agents can use their beliefs to keep track of the state of the dialogue, both the current dialogue state and the next dialogue state.

If a Speak action is selected, the dialogue manager will search for a dialogue in the dialogue pool with the corresponding attributes. For example: current state of ``Start'', a next state with ``Greeting'', and ``-, Rude'' as ``Meaning'' and ``Style'' parameters correspondingly. As shown in Table \ref{tab:dialogue} the dialogue manager would return the utterance ``Yo!?'' and ``What'up''. If multiple dialogues are available the system will chose one randomly. In this case the so called ``style'' flag is used, in this case, to influence the politeness level of the dialogues. FAtiMA-Tookit's authoring tool makes the process of creating and debugging dialogue relatively simple, however, naturally, complex scenarios require a lot manual labor both in terms of handling all the different states but also in giving more options to the users. 

\begin{table}
\caption{Example of different Dialogues defined by the Current State, the Next State, the Meaning and Style, usually used as auxiliary flags and the Utterance.}
  \begin{tabular}{ | c | c | c | c | c |}
  \toprule
    \textbf{Current State} & \textbf{Next State} & \textbf{Meaning} & \textbf{Style} & \textbf{Utterance} \\ 
    \midrule
    Start & Greeting & - & Rude & Yo!? \\ \hline
     Start & Greeting & - & Rude & What'up!? \\ \hline
    Start & Greeting & - & Polite & Hi, how are you? \\ \hline
    Greeting & GreetReply & -  & Polite & I'm great, how about you? \\ \hline
    Greeting & GreetReply & -  & Rude & Go away \\ \hline
    Greeting & GreetReply & - & - & Nothing special. \\ 
  \bottomrule
  \end{tabular}
  \label{tab:dialogue}
\end{table}

\subsection{Agent Modelling is an Iterative Process}

The nature of the process of creating a SIA interaction scenario makes it an iterative exercise. This means that like most types of user-based interaction experiences there is a lot of back a forth between different ``authoring phases''. In particular, going through the scenario as ``player'' is a reoccurring and important task. Is the interaction going through the right beats? Do the dialogues appear realistic? Are the agents properly generating emotions? 

\begin{figure}[ht]
\centering
\includegraphics[width=0.9\columnwidth]{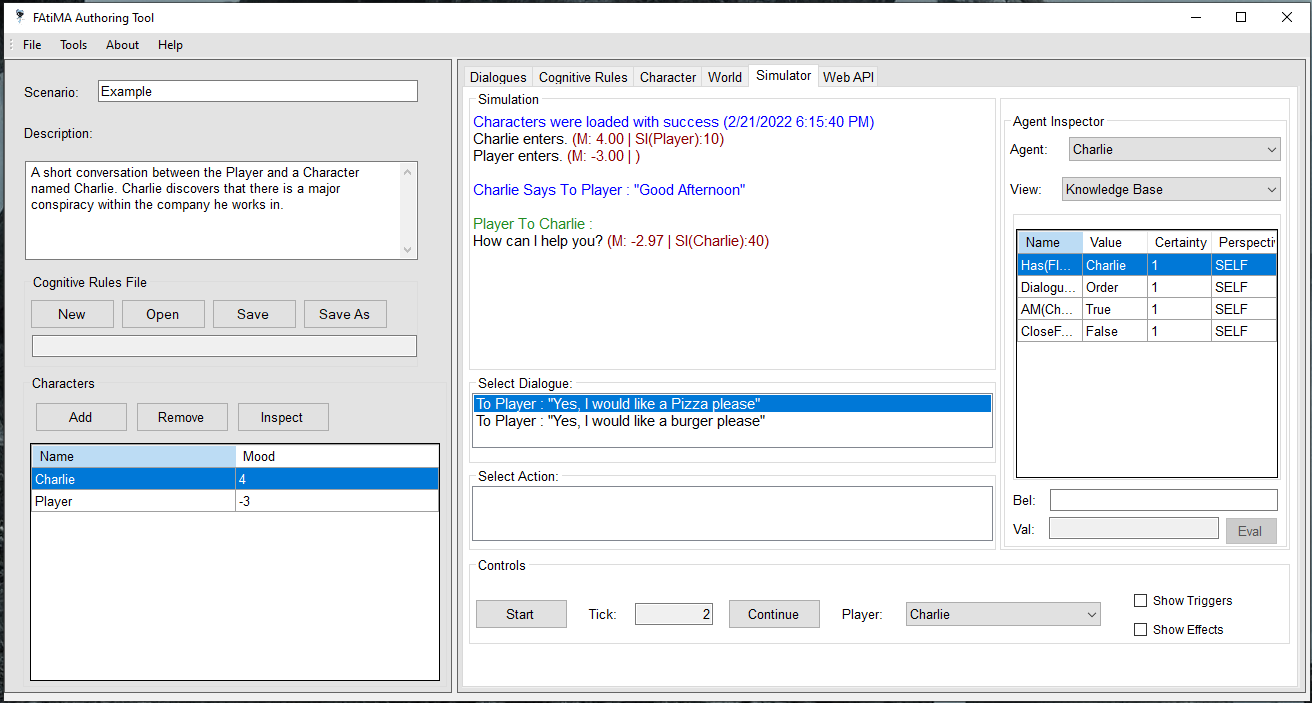} 
\caption{FAtiMA-Toolkit Authoring Tool's Simulator component which allows authors to quickly prototype and test a SIA-based scenario \cite{mascarenhas2021fatima}}
\label{fig:fatima}
\end{figure}

It is highly important to support and allow for authors to easily prototype and test the scenarios which they have created. Furthermore, play-testing the interaction leads to the inescapable bug detection, implementing bug-fixes and overall scenario refinement. A smooth and quick transition between these tasks is essential to the authoring experience. Each scenario component should be interchangeable and modifiable at any time, without much effort. One of the most important features of the authoring tools we have mentioned is exactly that. Figure \ref{fig:fatima} displays FAtiMA-Toolkit's Simulator components which allows developers to play as any of the agents and simulate their behaviour.

\section{Interpretability, Transparency and Complexity in FAtiMA Toolkit}
\label{sec:evaluation}

To explore to what degree a theory-driven authoring tool supports understandability during the authoring process, from the point of view of the user, we conducted two user-studies using FAtiMA-Toolkit \ cite{mascarenhas2021fatima}. 
One study analyses how interpretable and transparent the authoring steps are for experienced and inexperienced users (\textbf{RQ1}). The goal is to evaluate whether FAtiMA-toolkit and underlying meta-model is comprehensible and interpretable, from a human perspective, and inform the design of an \textit{Explainable Social Agent Authoring Tool}. The other study study focus on complexity of scenarios that an author is able to create and what artefacts they use to complete the task (\textbf{RQ2}). This analysis is based on the assumption that the amount of rules will certainly increase the quality a scenario, but will compromise interpretability. We investigate how to conceptualize complexity in this context and what is its impact on how the authoring tool is used. 




\subsection{Model Understandability}

Understandability is the core concept of Explainable AI (XAI), as it portrays the degree to which a human can understand a model and a decision made by it. This can be an objective and subjective measure depending on several factors, including how a human understands the concepts of the model. In this study, we intend to evaluate whether FAtiMA-toolkit is understandable from the point-of-view of a human, including if the meta-model is comprehensible and interpretable.

From a practical standpoint understandability is related to the ease of use, i.e. the effort required to read and interpret the meta-model. Previous research survey in \cite{houy2012understanding} have assessed this concept by looking into reaction times or response latency, correctness of response and ability to reason and substitute or add artefacts to guide the scenario to a certain output. 


\subsubsection{Procedure}
We conducted a between-subjects experiment with one treatment: knowledgeable users vs novice users. Our goal is to assess whether the effort required to read and correctly interpret FAtiMA-Toolkit constructs (meta-model), connections and artefacts depend on the knowledge of the participants. A variable that is identified as cornerstone to XAI and understandability in general \cite{arrieta2020explainable}.

Participants were recruited through email sent to students mailing lists. The goal was to obtain specialized subjects for this evaluation, i.e., individuals that have already created social scenarios using FAtiMA-toolkit and individuals that have the same academic background (Computer Science) but have not interacted with FAtiMA toolkit before.

The study was conducted online thus, a link was sent to participants which redirected them to an online survey, automatically assigning them a unique ID, preserving anonymity. The survey started with a description of the task, research goals and consent form. Upon acceptance of the conditions, the participants received the following instructions:
\begin{enumerate}
    \item Step 1. Watch the following tutorial video on “What are intelligent agents?”
    \item Step 2. Complete the following questionnaire divided into different tasks (see Appendix)
\end{enumerate}

The tutorial video consisted of a short 10 minute presentation on the basic concepts behind social intelligent agents. It included the perception-action cycle, the usual internal components of agents such as goals, beliefs and desires and a few examples of how agent modelling architecture define actions and emotions. Section \ref{sec:authoring} illustrates these steps in different modelling architectures. 

\subsubsection{Participants}
A total of 20 participants, 15 male and 5 female participated in the study. Each experimental group contained 10 different individuals. The first group, to whom we will name as the ``Experienced'' group consisted in Computer Science students which had previous authoring experienced with FAtiMA. Each individual had contributed to the creation of at least 2 different authoring scenarios, at least 10 hours of authoring experience and a 3 hour ``Getting Started'' workshop. The second group, the ``Inexperienced'' group was also only formed by Computer Science students but they had never heard or used FAtiMA-Toolkit or similar tools before.

The ``Experienced'' group averaged 23.8 years of age (SD=5.7) and, on average, spent around 23.34 minutes (SD=7.09) to complete the tasks.
The ``Inexperienced'' group averaged 22 years of age (SD=2.5) and  spent around 26.39 minutes (SD = 12.45) to complete the survey. All participants received a 20€ voucher by completing the survey. 

\subsubsection{Tasks}

Following the tutorial video, participants were presented with a set of 10 different questions. The question or task's order was fixed as they were designed to resemble the authoring steps in traditional agent modelling tools.

To investigate meta-model understandability the tasks were divided into comprehension tasks and problem solving tasks \cite{patig2008preparing}. Comprehension tasks addressed Syntactic and Semantic questions regarding the meta-model and Problem Solving Tasks require that participants apply the meta-model and simulate a possible output given a situation description, using the system artefacts. See Table \ref{tab:tasks} for a correspondence.

\begin{table}
\caption{Tasks Level and their definitions}
  \begin{tabular}{|p{0.25\linewidth}|p{0.45\linewidth}|p{0.2\linewidth}|}
  \toprule
    \textbf{Task Level} & \textbf{Definition} & \textbf{Task Number} \\
    \midrule
   Specification & Creation of model artefacts & 1, 3 \\ \hline
    Surface Level/ Syntactic & Constructs of the metamodel and their relationships &
2  \\ \hline
   Surface Level/ Semantic & Understanding of the contents described & 2  \\ \hline
   Problem Solving &
Participants are requested to determine whether and how certain information can be retrieved from an artifact created by applying the meta-model & 4 to 8 \\
  \bottomrule
  \end{tabular}
  \label{tab:tasks}
\end{table}

\subsubsection{Measures}

To test our hypothesis we used as an independent variable the level of experience and familiarity with FAtiMA-Toolkit. As dependent variables we took into account the score achieved by participants, the number of correct answers, the time taken to complete the survey and the response latency (a measure of cognitive difficulty).

To evaluate if there was a learning effect of 
the survey and the presentation the Perceived Competence subscale of the Intrinsic Motivation Inventory (IMI) questionnaire \cite{mcauley1989psychometric} was used. It measures the general subjective learning effectiveness of the experiment. In addition to this, participants were asked to rate each question according to their perception of how much effort they put into its completion and their perceived difficulty level.

\subsubsection{Results: IMI - Perceived Competence} 

The perceived competence of participants was captured using 6 different items for both the conditions, before and after the survey. The value for the Cronbach Alpha was as follows: for the experienced group, before the survey was 0.895 and afterwards it was 0.9. 
For the inexperienced group before the survey was 0.616, which made us ignore one of the questions, leading to a 0.76. After the survey, for the same group the Cronbach's Alpha value was 0.89. These values allowed us average all of the items composing the scale and perform sample tests to the whole group.  

Afterwards an Independent Sample Test was performed to the Perceived Competence of both groups, before the experiment, and a significant difference was found between inexperienced (M=4.03, SD=0.18) and experienced participants (M=4.78, SD=0.94); t(18)=-2.146, p = 0.046. 
Here, the Perceived Competence levels were affected by their experience (F(1,18)=4.6, p=0.046. However the effect is not very strong as indicated by its partial eta squared of 0.2.

After completing the survey while both groups increased their Perceived Competence levels, no significant difference was found between the inexperienced  (M=4.97, SD=0.25) and experienced (M=5.15, SD=0.29) participants.

\begin{figure}[h]
     \centering
     \includegraphics[width=0.7\linewidth]{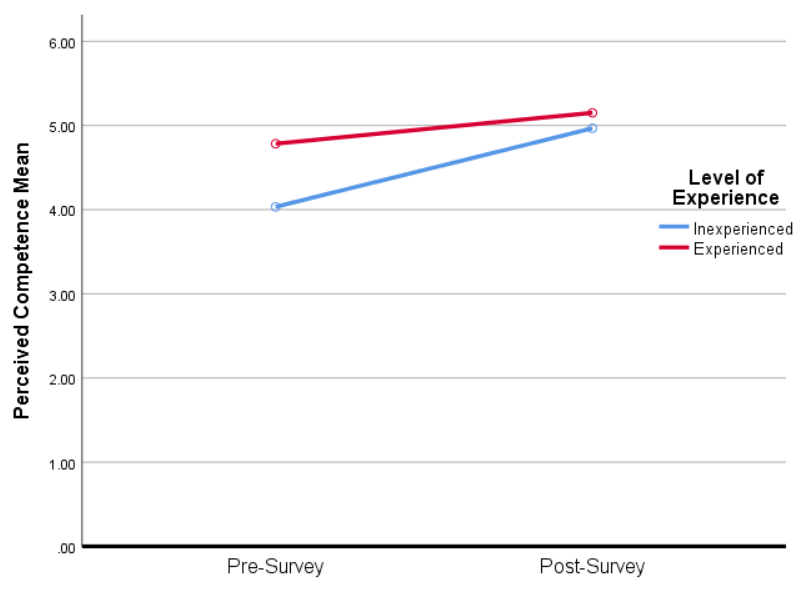}
     \caption{Difference between conditions regarding Perceived Competence before and after the experience}\label{fig:PerceivedCompetence}

\end{figure}

In terms of the impact of the experience within each groups, there was a significant difference in the scores for Perceived Competence for Inexperienced users, before the survey (M=4.03, SD=0.18) and after (M=4.97, SD=0.25); t(9)=-3.85, p = 0.004.
Additionally significant difference was also found in the scores for Perceived Competence for Experienced users, before (M=4.78, SD=0.3) and after (M=5.15, SD=0.29); t(9)=--2.8, p =0.021. Figures \ref{fig:PerceivedCompetence} capture these effects. 
\\


\subsubsection{Results: Task Success}

An independent-samples t-test was conducted to determine whether there is a difference between the overall scores of experienced and inexperienced participants. In total the experiment had 10 different questions graded equally, 0 in case it was the wrong answer and 1 if it was correct. Thus the maximum score attainable was 10. The results indicate a significant difference between inexperienced (M=7.7, SD=1.25) and experienced (M=9, SD=0.94) participants. [t(18)=-2.6, p=0.017]

Our focus however, is not in the overall difference between the groups but in which types of questions participants were able to successfully answer. Thus, our analysis must take into account the type of task and their objective as defined in Table \ref{tab:tasks} .

Regarding ``Specification'' level tasks, there was no difference between the groups, almost all Experienced and Inexperienced participants were able to successfully perform Task 1 and Task 3 (8 and 9 correct answers in both groups for each task, respectively). Task 2 also showed practically no difference between the conditions since, from each group, only one person failed to correctly complete the task.

In terms of tasks related to Problem Solving, differences between the groups were found. More specifically in Tasks 5, 6.1, 6.2 and 8.2. From these, the most significant one was Task 6 which concerned emotional appraisal. A significant different was found in Task 6.1 between inexperienced  (M=0.3, SD=0.48) and experienced (M=0.9, SD=0.32) participants [t(18)=-3.29, p=0.04]. It is important to notice that 0 value corresponds to an incorrect answers while 1 corresponds to a correct response. Furthermore in Task 6.2, while no significant difference was found, the inexperienced group of 10 participants only provided 5 correct answers while the experienced group provided 8. 

Finally it is important to add the fact there was no difference between the number of correct answers in Task 4, 7 and 8.1 where scores were quite high. Figure \ref{fig:correctanswers} captures the number of correct answers given in each task between the two groups and between different task levels.



\definecolor{bblue}{HTML}{4F81BD}
\definecolor{rred}{HTML}{C0504D}
\definecolor{ggreen}{HTML}{9BBB59}
\definecolor{ppurple}{HTML}{9F4C7C}
\definecolor{oorange}{HTML}{F3A751}

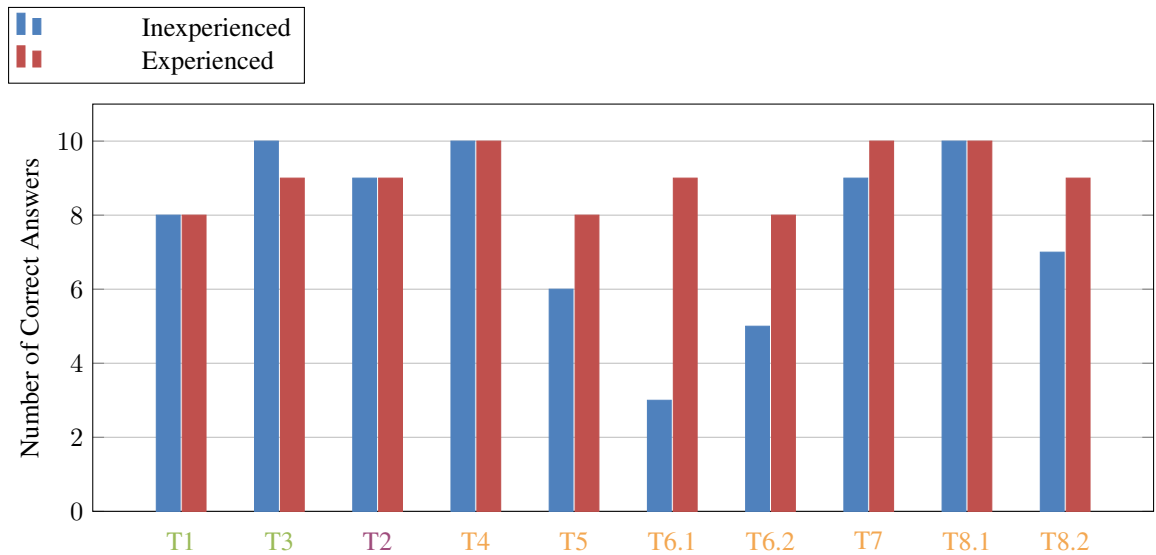
\begin{figure}
\begin{tikzpicture}
 \begin{axis}  
[     width  = 0.95*\textwidth,
        height = 7cm,
        major x tick style = transparent,
        ybar=2*\pgflinewidth,
        bar width=9pt,
        ymajorgrids = true,
        ylabel = {Number of Correct Answers},
        symbolic x coords={\color{ggreen}T1,\color{ggreen}T3,\color{ppurple}T2,\color{oorange}T4,\color{oorange}T5, \color{oorange}T6.1,\color{oorange}T6.2,\color{oorange}T7,\color{oorange}T8.1,\color{oorange}T8.2},
        xtick = data,
        scaled y ticks = false,
        ymin=0,
        legend cell align=left,
        legend style={
                at={(0.2,1.05)},
                anchor=south east,
                column sep=8ex}
    ]
 \addplot[style={bblue,fill=bblue,mark=none}] 
            coordinates {(\color{ggreen}T1,8)(\color{ppurple}T2,9)(\color{ggreen}T3, 10)(\color{oorange}T4,10)(\color{oorange}T5,6)(\color{oorange}T6.1,3)(\color{oorange}T6.2, 5)(\color{oorange}T7,9)(\color{oorange}T8.1,10)(\color{oorange}T8.2,7)}; 
 \addplot[style={rred,fill=rred,mark=none}]
            coordinates {(\color{ggreen}T1,8)(\color{ppurple}T2,9)(\color{ggreen}T3, 9)(\color{oorange}T4,10)(\color{oorange}T5,8)(\color{oorange}T6.1,9)(\color{oorange}T6.2, 8)(\color{oorange}T7,10)(\color{oorange}T8.1,10)(\color{oorange}T8.2,9)};  
  \legend{Inexperienced,Experienced}
\end{axis}  
\end{tikzpicture}  
\caption{Number of correct answers in each task by condition. Tasks are organized by their level: \color{ggreen}Specification, \color{ppurple}Surface Level:Syntactic and Semantic \color{black}and \color{oorange} Problem Solving.} \label{fig:correctanswers}
\end{figure}

\subsubsection{Results: Effort and Difficulty} 

During the survey participants were asked to rate each question according to their perception of how much effort they put into the task and how difficult they thought it was.  Users were rated the amount of effort using a Likert-Scale from 1-5 where 1 was ``No effort at all'' and 5 represented ``A Lot of effort'' . Regarding the difficulty, participants rated the questions using a 1-5 Likert Scale where 1 was ``Very Easy'' and 5 was ``Very Hard''.

An independent-samples t-test was conducted to determine whether there is a difference between the effort and difficulty  of experienced and inexperienced participants. No significant difference was found between the overall effort put into the tasks between inexperienced (M=2.44,SD=0.48) and experienced (M=2.27, SD=0.38) participants. In terms of difficulty however a significant difference was found between inexperienced (M=2.53, SD=0.09) and experienced (M=2.16, SD=0.36) participants [t(18)=2.58, p=0.019]

The difference between the difficulty and effort scores was uniform throughout most of the tasks regardless of their type. Nonetheless both experienced and inexperienced participants found Tasks 1, 2 and 3 to be relatively easy and had to put some effort into completing them. The highest difficult and effort results were found in Tasks 5, 6.1, 6.2 and 8.1, incidentally the exact ones where the scores were lower. Participants found the most difficult question to be 6.2 (M=3.25, SD=0.79) and also the one that required most effort  (M=3.05, SD=0.95) which equates to moderate difficulty and moderate effort.

\subsubsection{Results: Time and Response Latency}

An Independent Samples Test was conducted to analyze the difference of time of completion between groups (in minutes). 
No significant difference was found between inexperienced participants (M=26.39 ,SD=12.45) and experienced (M=23.34, SD=7.09) participants. Additionally, a Mann-Whitney-U was conducted to determine whether there is a difference between the time spent watching the presentation (in minutes) and no significant difference was found between inexperienced participants (M=9.05,SD=8.1) and experienced (M=6.37, SD=1.97) participants.

The final question of the survey asked participants to make an estimate of the time they used to complete the survey. Interestingly, both of the groups had similar estimates. Experienced participants averaged at 31.8 minutes (SD=9.95) while Inexperienced participants estimated they spent around 31.4 minutes (SD=7.2).

Regarding the time, it is important to note that while the platform used to host the experiment was able to keep track of the start and end time of the surveys, unfortunately there was a lot of outside variables we are not able to account for. Since the test was done remotely participants had around a week to complete and often started a task and only finished it long after (in one case even, days after). Thus, within this metrics there is a lot of disparaging results.

\subsubsection{Discussion}


Regarding \textbf{Research Question 1}, the study's results support the claim that meta model concepts, evaluated in the survey, are comprehensible and interpretable for experienced users and, in some degree, most are also understandable for inexperienced users.

Experienced and Inexperienced participants showed proficiency at performing tasks related with the creation of meta model artefacts (Specification), of the relationship between the different constructs within the meta model (Syntactic) and  at understanding the contents each of them tries capture (Semantic).

In terms of problem solving: using meta model concepts to complete different tasks related to the authoring of intelligent social agents, there were some significant differences between the two conditions. More specifically, in terms of the ideas behind emotional models it is clear experienced participants were more proficient at to performing affective tasks.

Regarding overall difficulty and effort, participants found the survey to be between Easy and Neural Difficulty and both groups put little to moderate effort into completing the tasks.
Naturally, higher difficulty was found in the tasks related to Problem Solving coincidentally the one's where most inexperienced users failed. Even so, experienced users also felt these were the hardest and the one's that required the most effort. These findings suggest that emotional concepts can be more complex than other but, with some interaction experience, can be just as easily understood and used. 


\subsection{Analysis: Domain Model}

The meta model of theory-driven agent modelling approaches defines a higher level formalism that allows the representation of social states (such as specific social relations) and the representation of social effects of actions.

The domain model, in turn, is the result of an authoring process that uses the meta model as a representational formalism to create particular instantiations of relations, actions, rules and constraints for a particular domain using a particular framework. 


In theory, each layer between the meta and the domain model adds noise to the equation and might lead to a less interpretable output. While this barrier unavoidable, agent modelling tools strive to make this transition as transparent as possible. The resulting scenarios' intricacy must also be taken into account since complexity is often, considered as a direct opponent to interpretability \ cite{buhrmester2021analysis}.

In order to answer \textbf{RQ2} in the following section we will present an analytical study focused on the domain model itself. In particular the complexity of different scenarios made by different authors but with the same goal, to create a realistic human-agent interaction.

\subsubsection{Background: Games AI Course}\hfill \break

FAtiMA-Toolkit has been used by the Artificial Intelligence in Games course of Instituto Superior T\'{e}cnico of University of Lisbon in the Computer Science Master's Degree. For the past 3 years over a hundred students have used the tool to create human-agent interaction scenarios \ cite{mascarenhas2021fatima}. The following analysis will focus on the work of students from 2021/2022 semester. For their final project, students were tasked with making full use FAtiMA-Toolkit's features to create two different social agent experiences. 


\subsubsection{Procedure}\hfill \break

To get acquainted with the tool, students had a 1.5 hour practical class focused on FAtiMA Toolkit on top of the additional theoretical class regarding Interactive Narratives. This class, while no focused on the tool itself, approached several important topics such as the perception-action cycle,  computational models of emotion and social agent architectures.   

This final project had two sequential authoring tasks where students had to create two different scenarios. The first scenario was the implementation of a small story using the Authoring Tool. The objective of this task was familiarize students with the tool, set a baseline for all the groups and served as a checkpoint for the teachers.
The following is the description of the scenario given to students:

\begin{quote}
    \textit{``Peter was hungry so he went to a restaurant. Once there he ordered a hotdog. The waiter told
him they only served hamburgers. Peter told the waiter that was okay as well. After a while the
waiter brought Peter his food. The hamburger was burnt to a crisp. Peter complained to the
waiter. The waiter told Peter they had no more hamburgers. Peter immediately left the
restaurant without paying.''}
\end{quote}

In the second scenario students were given almost complete freedom regarding their ``creations''.
The scenario had to fill a few requirements such as: using 2 different agents, having 15 different dialogue states and use at least 3 different reasoner components such as the Emotional Appraisal, Emotional Decision Making and World Model. Naturally students were encouraged (by having a better grade) to expand upon these requirements which they did. 

Following the theoretical and practical classes on the topic of Social Intelligent Agents and FAtiMA-Toolkit, groups were given 2 weeks to complete their projects. The Toolkit's official website has a wide array of documentation and tutorials which are directly available to anyone that seeks them.

\subsubsection{Participants}\hfill \break

35 students, forming 16 groups of 2-3 authors each (M=2.31, SD=0.7) were tasked with creating social interaction scenarios using an affective agent modelling tool: FAtiMA-Toolkit. All of the participants were attending the fourth year of University of Lisbon's Computer Science Master's degree.

\subsubsection{Objectives}\hfill \break

The structure of a FAtiMA-Toolkit authored scenario, like most traditional agent-modelling approaches, is a set of rules with conditions within different components.
Scenarios have a number of agents with different beliefs and goals, dialogues with different dialogue states and components with rules and conditions, similar to the content in Section \ref{sec:authoring}

All of the groups were able to complete their tasks with some level of success, by at least completing the minimum requirements. Naturally, even when given the same exact task, different authors will have different ways of implementing their solution and make use of tools available to them.

Our objective with this analysis is not to look at each scenario with an qualitative mind but, instead, to use the information to establish a baseline for complexity and amount of authorial effort regarding different types of scenarios.

\subsubsection{Metrics}\hfill \break

 In order to gauge the complexity of an agent-modelling scenario it is necessary to take into account both the complexity its scale. In order to measure this we used ``artefacts''. An artifact can be: a rule, a condition, a belief, a dialogue, it captures anything that was manually authored by the author.
 
The number of artefacts created by an author can be used to compare the relative complexity of different scenarios. We consider that one action rule for instance is one artefact and one action rule with one condition is 2 artefacts and so on, as shown in Table \ref{tab:artefactexample}. Even when there are no conditions an action rule for instance can be triggered and was manually authored, thus, we consider it to be the baseline and one artefact. 
 
 \begin{table}
  \caption{Artefact Example: A scenario with two rules with 1 condition between them has the same complexity of a scenario with one rule with two conditions}
  \begin{tabular}{  c | c | c }  
    \toprule
   \textbf{Number of Rules} & \textbf{Number of Conditions} & \textbf{Number of Artefacts } \\
    \midrule
    1 & 0 & 1 \\ \hline
     1 & 1 & 2 \\ \hline
     1 & 2 & 3 \\ \hline
      2 & 1 & 3 \\ \hline
     4 & 4 & 8 \\
    \bottomrule
  \end{tabular}
  \label{tab:artefactexample}
\end{table}

It is possible to use the same logic to the other components. One agent is an artefact, a belief is an artefact, a goal is an artefact and a dialogue is an artefact. This, relatively straightforward metric allows us to, not only compare the overall complexity of scenarios but also  understand where students spent most of their effort.

In addition to this, in FAtiMA, time frames can be represented by \textbf{cycles}. The algorithm gives the chance for agents to perform an action. Thus, in the same cycle multiple agents can execute an action. In a typical human-agent interaction, for example, the agent would greet the player and the player would greet an agent. Both of these events occurred in cycle 1, for instance.

Using the ``Simulator'' tool \ cite{mascarenhas2021fatima} within FAtiMA-Toolkit it is possible roughly estimate how many cycles occurred in each scenario.  Naturally, it is frequent for the author to design multiple endings according to the user«s options which, in turn, might increase or decrease both the space and the length of the interaction experience. Thus the ``cycle'' can only be used as a rough estimate.


\subsubsection{Task 1 Results}\hfill \break

Regarding Task 1, with 16 different participants (scenarios), to implement the previously mentioned description, on average 95.6 artefacts (SD=36.7) were created. The mean amount of different dialogues was 18.2 (SD=14.7) with 2 different agents (M=2.1, SD=0.250). Finally, regarding the interaction itself the average number of cycles was 7.2 (SD=1.95) with almost half of the scenarios having more than one ending (M=1.4, SD=0.5). 

In addition to this we asked groups to estimate the time spent learning the framework and then implementing the scenario itself. Not all of the groups responded, however, from a sample of 14 groups, each member spent, on average 2 hours (SD=0.68) learning about the toolkit and then the whole group spent around 4 hours (SD=1.0) implementing the story. 

\begin{figure}
\begin{tikzpicture}  
  
\begin{axis}  
[    width  = 0.95*\textwidth,
        height = 7cm,
        major x tick style = transparent,
        ybar=2*\pgflinewidth,
        bar width=9pt,
        ymajorgrids = true,
        tick label style={font=\footnotesize} ,
    ylabel={Average Artifact Count}, 
    symbolic x coords={Internal State, Dialogues, EmotionAppraisal, DecisionMaking, World Model}, 
    xtick=data,  
     nodes near coords, 
    nodes near coords align={vertical},  
    ]  
\addplot[style={bblue,fill=bblue,mark=none}] coordinates {(Internal State,12.3) (Dialogues,18.1) (EmotionAppraisal,4.6) (DecisionMaking,41.3) (World Model,19) };  
  
\end{axis}  
\end{tikzpicture}  
  \caption{Task 1 Results: Average Artifact Count by Component}
    \label{fig:task2results}
\end{figure}
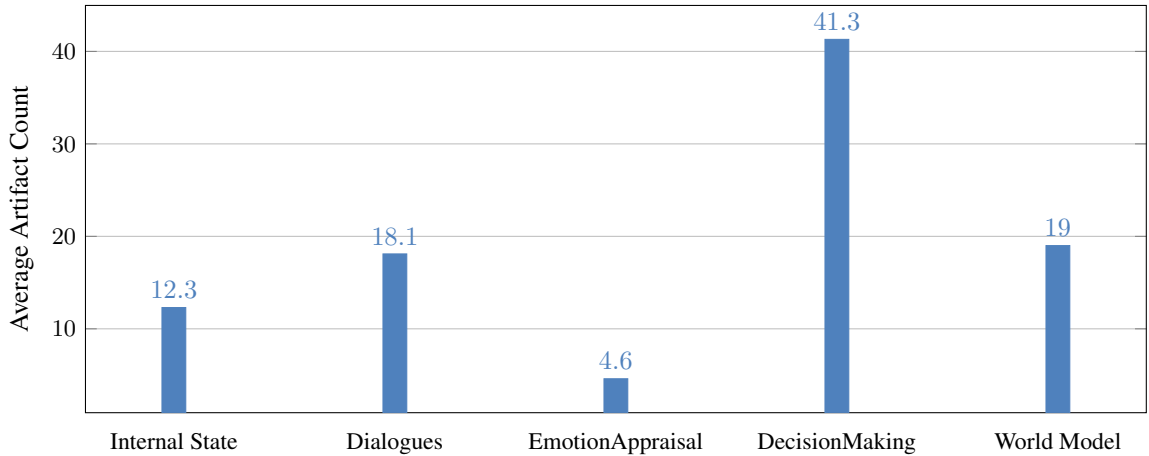

\subsubsection{Task 2 Results}\hfill \break

Regarding Task 2, it's open ended nature led to a  wide range of different scenarios such as a blind date, an AA meeting, a blackjack game, a stand-up comedy show among others. Overall, the number of artefacts created on average was 178.8 (SD=101.8), the number of dialogues 57.4 (SD = 37.6) with the average number of agents being 3 (SD=1.2). The average scenarios had more than 3 endings (M=3.44, SD=1.54) and  the average number of cycles was 15.38 (SD=12.06).

Using the same methodology as Task 1, we asked groups to estimate the time spent learning the framework and then implementing the scenario itself. While not all of the groups provided us with an answer, from a sample of 11 groups, each member estimated they spent, on average 5 hours (SD=1.6) learning about the toolkit and then the whole group spent around 8 hours (SD=5.9) implementing their idealized scenario. 

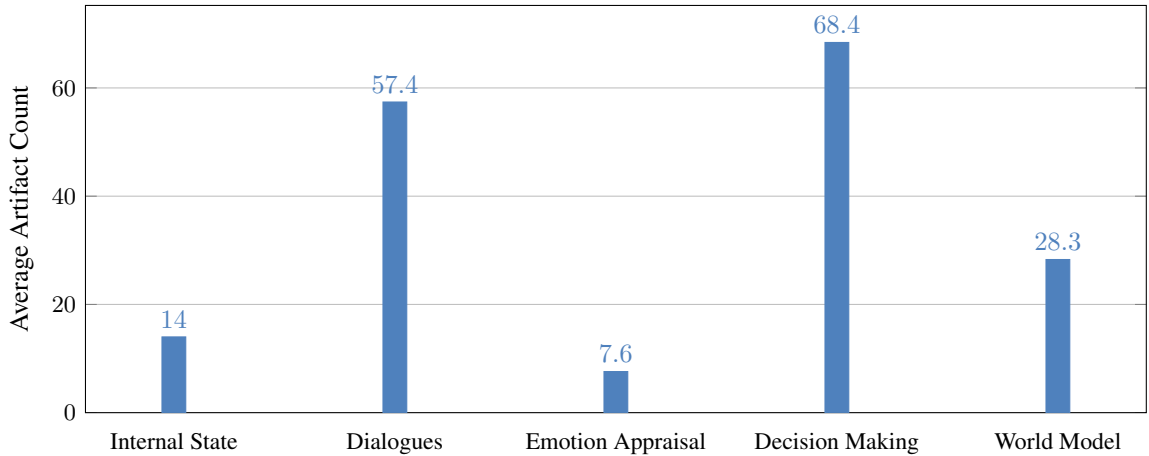
\begin{figure}
  
\begin{tikzpicture}  
  
\begin{axis}  
    [     width  = 0.95*\textwidth,
        height = 7cm,
        major x tick style = transparent,
        ybar=2*\pgflinewidth,
        bar width=9pt,
        ymajorgrids = true,
        tick label style={font=\footnotesize	} ,
        ylabel = {Average Artifact Count},
         symbolic x coords={Internal State, Dialogues, Emotion Appraisal, Decision Making, World Model},
        xtick = data,
        scaled y ticks = false,
        ymin=0,
        nodes near coords,
        legend cell align=left,
        legend style={
                at={(0.2,1.05)},
                anchor=south east,
                column sep=8ex
        }
    ]
\addplot[style={bblue,fill=bblue,mark=none}] coordinates {(Internal State,14) (Dialogues,57.4) (Emotion Appraisal,7.6) (Decision Making,68.4) (World Model,28.3)};  
  
\end{axis}  
\end{tikzpicture}  
  \caption{Task 2 Results: Average Artifact Count by Component}
    \label{fig:task2results}
\end{figure}

\subsubsection{Discussion}\hfill

Students spent most of their time and effort in creating dialogues and decision making related artefacts. There's a number of possible reasons that can explain this, starting with the objective of their task. When asked to create any type of scenario, most opted into a conversational one and spent most of their time authoring the dialogue and dialogue actions themselves. 

In addition to this, it is important to note that due to FAtiMA-Toolkit's hybrid approach, as described in Section \ref{sec:authoring}, the Decision Making and Dialogue Manager components are intrinsically connected, hence, we believe, the high number of Decision Making and Dialogue artefacts.

Notably it is possible to use the data and  establish an average baseline for authors. In order to create an interactive scenario with 3 agents, around 178 artefacts, 15 interaction cycles and at least 3 different endings one person would have to spend around 21 hours (8 * 2.3 working hours + 5 hours for learning) their time.  


\subsection{Overall Discussion}

The studies described above provides us with new and interesting insights into the understandability of theory-driven social agent's meta model and author's tendencies when creating domain models, in particular, using FAtiMA-Toolkit. 

Regarding \textbf{Research Question 1}, it is clear most concepts, with the exception of emotions, are easy to understand for both experienced and inexperienced users. 
After watching a 10 minute presentation participants with no previous experience in modelling intelligent social agents were capable of creating meta model artefacts, understanding the relationship between them and successfully performing problem solving tasks almost as well as experienced users.


In terms of \textbf{Research Question 2}, when University of Lisbon students had to create their own interaction experiences using FAtiMA-Toolkit, most students prioritized the dialogue and decision making components. The analysis allowed us to establish a baseline where for one person to create a scenario composed by 3 different agents, multiple endings and almost 200 artefacts had to spend around 21 hours. Notably, once again, the emotional components were left mostly unused when compared to the others.

Both studies revealed an apparent lack of comprehension and use of the affective components of traditional social agent frameworks. These findings can be explained by multiple factors. It is possible that, due to the lack of understanding and the usual complexity of the model behind the generation of emotions, authors avoid the use of such components. It is also possible that, since authors were tasked with creating a human-agent interactive scenario they put most of their work into the flow of conversation itself leaving other components such as emotional appraisal behind. 

In addition to this, while the use of affective elements significantly improve the impact and believability of social agents, perhaps traditional agent modelling tools do not motivate authors enough to justify their use. To simply put it, if there was a representation of the facial expression of the agents during the interaction, or a non-verbal behaviour simulator, authors could perceive, comprehend and be more motivated to use these systems better. Naturally, further studies need to be completed to confirm the explanation behind the results found.

\section{Enhancing explainability in existing Social Agent Architectures}
\label{sec:suggestions}

The work described in this paper provides key insights into some of the fundamental issues of creating socially intelligent agents using agent modelling tools. As we've seen, there is a wide variety of state-of-the-art approaches but none without its flaws. In our view, it is clear that in order to move towards the next generation of social agent frameworks, designers of such tools, should focus on the authoring experience itself, in particular, their explainability.

In order to create \textit{Explainable Social Agent Authoring Tools} developers should focus on improving the key elements we highlighted throughout this work:
\begin{itemize}
    \item \textbf{Interpretability}: the ability to demonstrate that a given input will produce a specific output in human-understandable way.
    \item \textbf{Transparency}: the frameworks inner workings should be readable by a human either through by its design itself or using post-hoc explainability techniques \ cite{arrieta2020explainable}. 
    \item \textbf{Comprehensibility}: associated to the model's output and artefact complexity along its ability to express information in natural language.
    \item \textbf{Control}: the creation of intelligent social agent interactions is an iterative process. It is highly important agent modelling tools provide authors, not only full control of every single component, but also the ability to easily test, change and refine their scenarios.
    \item \textbf{Scalability}: captures the amount of effort, manual labor, required to create a human-agent scenario. 
    \item \textbf{Interactiviy}: we believe it is of utmost important to undersline that the way the author and the tool communicate should help the designer understand what comes next in the authoring workflow, and help them achieve their authoring goals.
\end{itemize}

Our stance is that using a combination of: Theory, Data and Ontology-based approaches, it is possible to mitigate each others' shortcomings and provide a clearer and approachable authoring experience. The flexible and high precision attributes of a rule-based system will be used to extract SIA-related concepts from  data gathered by data-driven techniques, when applied to a social interaction data-set. Additionally the use of an ontology-based layer of representation will help the transition between the output of the data-oriented pipeline to the agent-modelling tool.

We envision the culmination of this hybrid approach as an authoring assistant that could directly support existing theory 
-driven frameworks (e. g. FAtiMA). The assistant would directly promote \textbf{Interactiviy} between the author and the tool itself. Moreover, operating within the framework with the objective of positively impacting the metrics mentioned above, it could be the bridge necessary to create an \textit{ Explainable Social Agent Authoring Tool}. 

For example, leveraging Ontology and Data-driven approaches the assistant could provide authors with an automated process that would take as input natural language descriptions or stories written by the author and, using social interaction corpora  extract and create a draft of the intended scenario. This ``skeleton'' would contain intelligent agents with context-based beliefs and goals, actions, emotional appraisal rules and even action effects. This feature would positively impact the tool's \textbf{Scalability} without sacrificing its \textbf{Interpretability} or \textbf{Comprehensibility}.

The authoring assistant would also be able to support post-hoc explainability techniques when needed. For instance in the case of the Emotional Appraisal, where it is clear further exposition is needed, it could support a direct simulator for authors to use, contributing to its \textbf{Transparency}. Furthermore the domain model study allowed us to establish baseline regarding the number of artefacts and complexity for each component. When authors are using the tool, the assistant could compare the current version of the scenario and advise users regarding improving and completing different sets, providing a clearer overall picture of the experience and increasing the authorial \textbf{Control} over the scenario.

\section{Conclusion}
Amongst the most promising applications of SIAs are serious games and social skills training environments. Their goal is to provide learners with realistic, but safe spaces enabling them to train specific verbal and nonverbal behaviors in order to adapt to socially challenging situations \cite{bosman2018virtual}. 

In this work the concept of \textit{Explainable Agent Authoring Tools} was introduced and framed within XAI literature.  
Afterwards an analysis of state-of-the-art social agent frameworks under this new perspective was performed. Its outcome was the identification and definition of 6 different important forces that should guide XSATs design: interpretability, transparency, comprehensibility, control, scalability and interactivity.  

Our analysis also showed that Theory-driven architectures are assumed to be understandable based on the fact that they are built upon well established psychological theories. In order to test this claim we set two different research questions focused on, first, the understandability of the theories behind the frameworks (meta-model) and second, the comprehensibility of the frameworks' generated output. 

Before presenting our answers and with the goal of providing a clearer picture over the authoring process, we presented how different theory-driven approaches represent similar concepts such as beliefs, actions, emotions and dialogues.

Using FAtiMA-Toolkit as a basis we performed 2 different studies, one focused on the meta model and the other on the domain model. The results presented here provide key insights into the use of such tools and the perception by authors of social agent concepts. It is clear some concepts are easier to understand and use for both experienced and inexperienced participants. An issue pertaining to the use and interpretability of computational models of emotions was found, which indicates further work regarding the understandability of emotional components is needed.


Finally, drawing from data presented in this work we set forth key concepts that can contribute and lead to the creation \textit{Explainable Agent Authoring Tools} 
The questions addressed in this work are essential to move forward to the next generation of agent modelling tools. We challenge designers of agent modelling tools to focus on measuring and improving their tools' explainability.

\section{Acknowledgements}
This study received Portuguese national funds from FCT - Foundation for Science and Technology through project: UIDB/50021/2020 and UIDB/04326/2020 and SLICE PTDC/CCI-COM/30787/2017.


\bibliographystyle{unsrt}  
\bibliography{references}  





\end{document}